\documentclass[prb,twocolumn]{revtex4-1}

\usepackage{amsmath}
\usepackage{graphicx}
\usepackage{epstopdf}
\usepackage{color}
\usepackage{url}
\usepackage{float}

\begin{document}

\title{Magnetically-induced phonon splitting in ACr$_2$O$_4$ spinels from first principles}

\author{Aleksander L. Wysocki}\email{awysocki@ameslab.gov}
\affiliation{Ames Laboratory, U.S. Department of Energy, Ames, Iowa 50011, USA}
\author{Turan Birol}
\affiliation{Department of Physics and Astronomy, Rutgers University, Piscataway, USA}

\date{\today}

\begin{abstract}
We study the magnetically-induced phonon splitting in cubic ACr$_2$O$_4$ (A=Mg, Zn, Cd, Hg) spinels from first principles, and demonstrate that the sign of the splitting, which is experimentally observed to be opposite in CdCr$_2$O$_4$ compared to ZnCr$_2$O$_4$ and MgCr$_2$O$_4$, is determined solely by the particular magnetic ordering pattern observed in these compounds. We further show that this interaction between magnetism and phonon frequencies can be fully described by the previously proposed spin-phonon coupling model that includes only the nearest neighbor exchange. Finally, using this model with materials specific parameters calculated from first principles, we provide additional insights into the physics of spin-phonon coupling in this intriguing family of compounds. 
\end{abstract}

\maketitle

\section{Introduction}

The interplay of spin and lattice degrees of freedom can lead to a variety of fundamentally and technologically interesting phenomena including the spin-Jahn-Teller effect in frustrated magnets, \cite{Yamashita,Tchernyshyov} magnetocapacitance, \cite{Kimura} and the linear magnetoelectric effect. \cite{Fiebig, Birol2012} One signature of this interplay is the influence of magnetic order on the vibrational spectrum of a material. In many transition metal oxides the spin correlations shift the phonon frequencies, and lead to the so called magnetodielectric effect.\cite{magnetodielectric_lawes, magnetodielectric_birol} Furthermore, if the long-range antiferromagnetic (AFM) order reduces the crystal symmetry, the onset of antiferromagnetism can result in a substantial splitting of phonon frequencies that are degenerate in the paramagnetic (PM) phase, even when the change in the crystal structure is undetectable.\cite{Massida,Luo, Sushkov, Fennie, Chan07} This phonon anisotropy is a non-relativistic effect which originates from the changes in hybridization due to spin ordering. In particular, the phonon splitting can be phenomenologically explained by a dependence of the exchange interactions on the atomic positions. \cite{Baltensperger,Baltensperger2}

Chromium spinels ACr$_2$O$_4$ (A=Mg, Zn, Cd) are a particularly interesting class of frustrated antiferromagnets that exhibit strong spin-phonon coupling. In the PM phase, group theory predicts, and experiments confirm, the presence of four triply degenerate infrared (IR)-active phonon modes. Below the  N\'eel temperature, however, one of these phonon modes undergoes a large splitting into a singlet and a doublet \cite{Sushkov,Aguilar,Kant,Kant2}. This feature and its magnitude  in  was argued to be a consequence of a dominant role of the nearest-neighbor (nn) direct Cr-Cr exchange interaction~\cite{Sushkov}. Fennie and Rabe developed in Ref. [\onlinecite{Fennie}] a general approach to incorporate material specific information from first principles into spin-phonon coupling models. They demonstrated that the spin-phonon coupling model with only nearest-neighbor (nn) exchange interaction and parameters derived from first principles provide a full description of experimentally observed magnetically-induced phonon splitting in ZnCr$_2$O$_4$.\cite{Fennie} This model has later been successfully applied to many other systems as well.\cite{Lee,Kumar,Zhou2013, Shen2008}

Magnetically-induced phonon splitting have been also observed in other Cr-based spinels including MgCr$_2$O$_4$ and  CdCr$_2$O$_4$. Interestingly, the sign of the phonon slipping observed for MgCr$_2$O$_4$ and ZnCr$_2$O$_4$ ($\omega_{\rm singlet}$ $>$ $\omega_{\rm doublet}$) \cite{Sushkov,Kant2} is opposite to that observed in CdCr$_2$O$_4$ ($\omega_{\rm doublet}$ $>$ $\omega_{\rm singlet}$). \cite{Aguilar,Kant} In all three of these compounds, the sign of the nn exchange interaction is the same, however its magnitude compared to further neighbour exchanges is dramatically different. In the Mg and Zn compounds the  nn exchange interactions  are two orders of magnitude larger than all other exchange interactions, while in CdCr$_2$O$_4$  the nn interaction is the same order as the second neighbor interaction. Based on this fact Kant \emph{et al.} \cite{Kant2} concluded that the spin-phonon coupling model with only nn exchange interaction cannot explain the magnetically induced phonon anisotropy in ACr$_2$O$_4$ spinels, and instead proposed that the phonon splitting is generally controlled by a nondominant, next neighbour exchange interaction.

In this paper we use first principles calculations to study the magnetically induced phonon anisotropy of the zone-center polar modes in  ACr$_2$O$_4$ (A=Mg, Zn, Cd, Hg) spinels. We show that the different magnetic orderings characteristic for these spinels lead to different signs of the phonon splitting. In particular, we explain the opposite sign observed for ZnCr$_2$O$_4$ and MgCr$_2$O$_4$ compared with CdCr$_2$O$_4$ which have distinct magnetic ground states. We find that the spin-phonon coupling model of Ref. [\onlinecite{Fennie}] with only the nn exchange interactions can very accurately describe \emph{ab initio} values of phonon frequencies for all the spinel compounds we considered.

\section{Methods}

The first principles calculations were performed using the density functional theory (DFT) within the rotationally invariant DFT+U method \cite{Liechtenstein} and the PBEsol approximation to the exchange-correlation functional \cite{PBEsol}. Similarly as in Ref. \onlinecite{Fennie} we used U = 3 eV and J = 0.9 eV, the parameters that accurately reproduce photoemission spectra and band gaps in sulfur Cr spinels \cite{Fennie2}.

The Kohn-Sham equations were solved using the projector augmented wave method \cite{Blochl} as implemented in the VASP code \cite{Kresse,Kresse2} with the plane wave cutoff of 500 eV and 6$\times$6$\times$6 $\Gamma$-centered k-point mesh in the primitive unit cell for the cubic structure (for larger cells the k-point mesh was scaled accordingly).
Phonon frequencies and eigendisplacements were calculated using the frozen phonons method using symmetry adapted modes obtained from the ISOTROPY package \cite{Isotropy}.
The spin-orbit coupling was neglected in the calculations. Structural relaxations were performed in the ferromagnetic (FM) state that preserves the cubic symmetry.

\section{Crystal and magnetic structure}

At high temperatures ACr$_2$O$_4$ spinels have a cubic ($Fd\bar3m$) structure where A$^{2+}$ ions are in tetrahedral oxygen environment and form the diamond lattice, while Cr$^{3+}$ ions are surrounded by octahedral oxygen cages and form a pyrochlore lattice, see Fig. \ref{spinel}a. The calculated structural parameters are shown in supplementary materials \cite{SM}.

\begin{figure}
\begin{center}
\includegraphics[width=1.0\hsize]{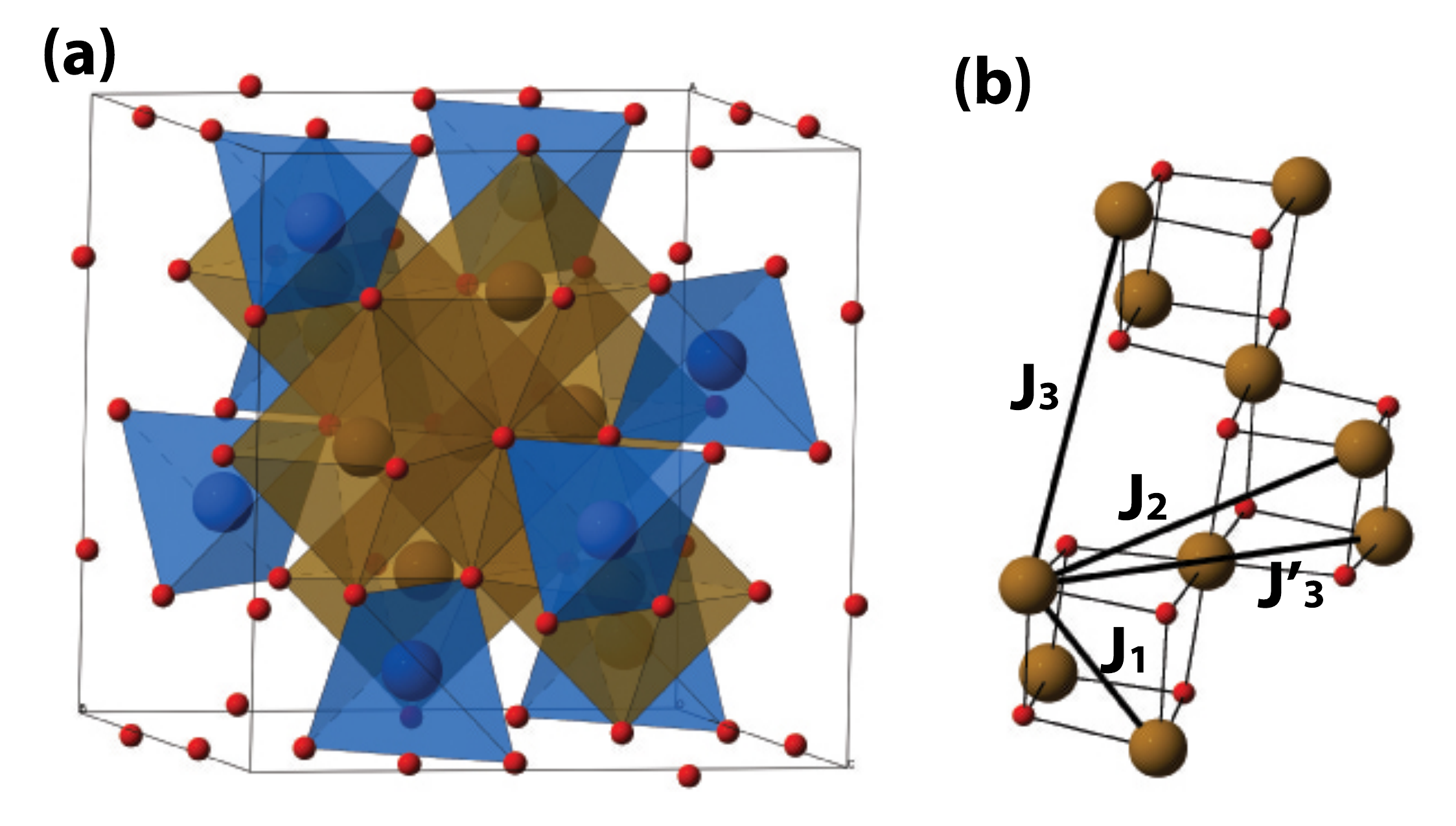}
\end{center}
\caption{(a) Cubic crystal structure of ACr$_2$O$_4$ spinels consisting of Cr-centered octahedra and A-centered tetrahedra. (b) Magnetic exchange couplings up to the third Cr neighbors; note that there are two nonequivalent types of third neighbors which have distinct exchange parameters: $J_3$ and $J_3'$. Brown, blue and red spheres denote Cr, A and O atoms, respectively.}
\label{spinel}
\end{figure}

The octahedral crystal field splits the Cr 3$d$ orbitals into a lower-lying $t_{2g}$ triplet and a higher-energy $e_g$ doublet. Cr$^{3+}$ has three outer electrons that fill the majority $t_{2g}$ states which results in a net Cr spin $S$=3/2. We found the exchange interaction parameters between Cr spins by fitting \emph{ab initio} energies of different collinear magnetic configurations to the Heisenberg Hamiltonian

\begin{equation}
H=\sum_{ij}J_{ij}\mathbf{S}_i\cdot\mathbf{S}_j
\end{equation}

where the summation is over Cr ions, $J_{ij}$ are the exchange parameters, and $\mathbf{S}_i$ is the unit vector indicating the direction of the spin at Cr site $i$. We considered exchange parameters up to the third neighbors (see Fig. \ref{spinel}b) as further neighbors are known to have negligible exchange couplings \cite{Yaresco}. The results are presented in Table \ref{exchanges}.

\begin{table}
\centering{}
\begin{tabular}{r|cccc|cccc|cccc}
\hline
\hline
 A & $J_1$ &$J_2$ &$J_3$ &$J_3'$ & \multicolumn{4}{|c} {$\mathcal{J}_{\lambda\perp}/\omega^{\text{PM}}_{\lambda}$} & \multicolumn{4}{|c} {$\mathcal{J}_{\lambda\parallel}/\omega^{\text{PM}}_{\lambda}$}\\
\cline{6-13}
    &           &         &         &        & 1 & 2 & 3 & 4 & 1 & 2 & 3 & 4\\
\hline
Mg & 3.81&-0.07&0.05&0.18 & 4.5 & 2.4 & 0.4 & 0.3 & -0.1 & -0.1 & -0.7 & -0.7 \\
Zn &  3.81&-0.08&0.14&0.16 & 2.0 & 5.0 & 0.8 & 0.2 & -0.0 & -0.1 & -0.8 & -0.7 \\
Cd &  0.29&-0.10&0.08&0.15 & 1.3 & 3.3 & 0.7 & 0.7 & -0.0 & -0.0 & -0.5 & -0.2 \\
Hg &-0.58&-0.00&0.21&0.13 & 0.8 & 3.2 & 0.9 & 0.6 & -0.0 & -0.0 & -0.5 & -0.1 \\
\hline
\hline
\end{tabular}
\caption{The exchange couplings (in meV) and $\mathcal{J}_{\lambda\perp,\parallel}/\omega^{\text{PM}}_{\lambda}$ parameters (in cm$^{-1}$) calculated for different ACr$_2$O$_4$ spinels. }
\label{exchanges}
\end{table}

The nn exchange parameter, $J_1$, is a dominant interaction for all compounds. This coupling arises from the competition between AFM direct exchange and FM 90$^{\circ}$ superexchange \cite{Goodenough}. For A ions with small ionic radii, like Mg or Zn, the direct exchange mechanism dominates resulting in a strong AFM $J_1$. However, for larger A ions the lattice parameter and the nn Cr-Cr distance increases \cite{SM} which diminishes the direct exchange contribution. In particular, for CdCr$_2$O$_4$ the AFM $J_1$ is reduced by an order of magnitude while for HgCr$_2$O$_4$ the superexchange contribution overcomes the direct exchange resulting in a (small) FM $J_1$.

Exchange couplings beyond nn originate from higher-order superexchange processes \cite{Dwight} and, in general, are smaller than $J_1$. However, while for MgCr$_2$O$_4$ and ZnCr$_2$O$_4$ these interactions are negligible compared to the nn exchange, for CdCr$_2$O$_4$ and HgCr$_2$O$_4$ compounds the $J_3$ and $J_3'$ exchanges become relevant.

The AFM nn exchange interaction is frustrated on the pyrochlore lattice since spins forming a tetrahedron cannot be all antiparallel to each other. The energy due to $J_1$ is minimized when for all tetrahedra the total spin is zero, i.e. in each tetrahedron two spins are parallel while the other two point in the opposite direction. There are, however, many such two-up-two-down configurations which can be different in different tetrahedra leading to infinite degeneracy. Consequently, the magnetic ground state is determined by further exchange couplings \cite{Chern}, magnetoelastic effects \cite{Yamashita,Tchernyshyov} or relativistic interactions \cite{Chern2} leading to complicated, often noncollinear, orderings.

\begin{figure}
\begin{center}
\includegraphics[width=1.0\hsize]{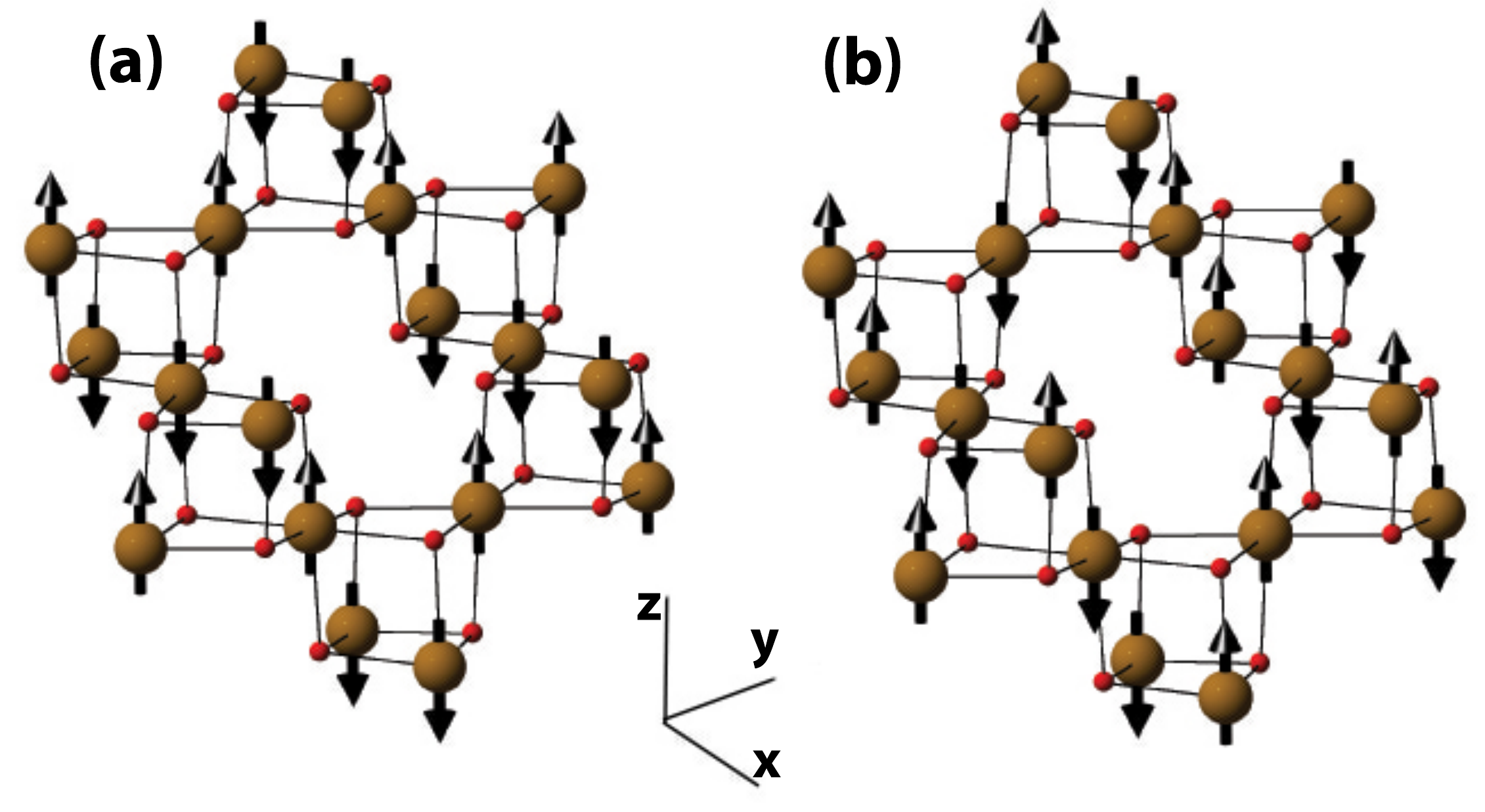}
\end{center}
\caption{Two collinear magnetic orderings relevant for ACr$_2$O$_4$ spinels. (a) AFM-I order that is similar to the true magnetic ground state in ZnCr$_2$O$_4$ and MgCr$_2$O$_4$. (b) AFM-II order that approximates the spin order in CdCr$_2$O$_4$. Brown and red spheres denote Cr and O atoms, respectively.}
\label{orderings}
\end{figure}

There are two primary collinear magnetic orders that are relevant for ACr$_2$O$_4$ spinels. These are shown in Fig. \ref{orderings} and we denote them as AFM-I and AFM-II. The AFM-I ordering is similar to the true magnetic ground state in ZnCr$_2$O$_4$ \cite{Tchernyshyov} and MgCr$_2$O$_4$ \cite{Xiang} while the AFM-II approximates the spin order in CdCr$_2$O$_4$ \cite{Chern}. Both spin orderings satisfy the two-up-two-down rule in each tetrahedron but they differ in relative orientations of spins in neighboring tetrahedra. In particular, for AFM-I the nearest neighbors in the $xy$ plane are parallel while for AFM-II they are antiparallel. As we will see below, this difference has a profound effect on the magnetically induced phonon anisotropy.

\section{Phonon frequencies}

We now focus on the influence of magnetic order on the zone-center polar phonons. In addition to the AFM-I and AFM-II orderings which are relevant for this class of compounds (see above), we also considered the FM order since which has the same cubic symmetry as the PM state.

In the FM state there are four triple degenerate polar phonon modes, each transforming according to the $T_{1u}$ irreducible representation of the $O_h$ cubic point group. In order to compute the corresponding phonon frequencies we considered symmetry-adapted $T_{1u}$ modes, $f_{n,\alpha}$. Here $n=1,2,3,4,5$ is the mode number (in addition to four polar modes we need to include the acoustic mode that also has a $T_{1u}$ symmetry) and $\alpha=x,y,z$ labels the row of $T_{1u}$ such that $f_{n,\alpha}$ transforms as a vector along the $\alpha$ axis. The symmetry-adapted modes, $f_{1,\alpha}$, $f_{2,\alpha}$, and $f_{4,\alpha}$, involve atomic displacements along $\alpha$ of the entire A, Cr, and O sublattice, respectively. On the other hand, atomic displacements associated with $f_{3,\alpha}$ and $f_{5,\alpha}$ take place in the plane perpendicular to the $\alpha$ axis and involve chromium and oxygen atoms, respectively (see the supplementary materials for more details).

\begin{figure}
\begin{center}
\includegraphics[width=1.0\hsize]{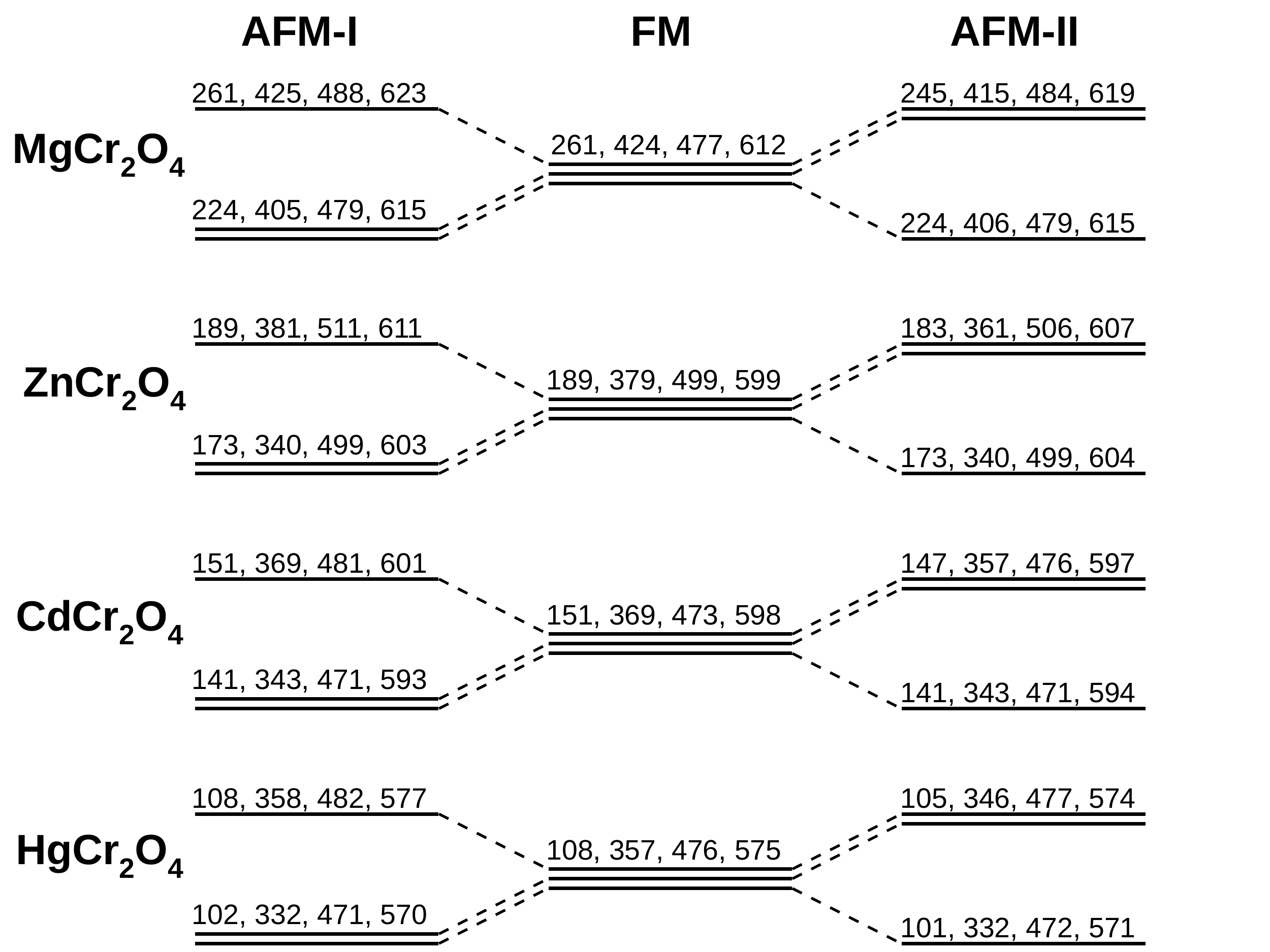}
\end{center}
\caption{Calculated $T_{1u}$ phonon frequencies (in cm$^{-1}$) of different ACr$_2$O$_4$ spinels for different magnetic orderings. In the FM state (middle) there are four optical $T_{1u}$ modes and each one is three-fold degenerate. In the AFM-I (left) and AFM-II (right) states each triplet splits into a singlet and a doublet. For the AFM-I (AFM-II) ordering the singlet (doublet) has higher frequency for all compounds and for all phonon modes.}
\label{phonons}
\end{figure}

Condensing the symmetry-adapted modes for a given $\alpha$ and evaluating the Hellman-Feynman forces for the FM state, we constructed the 5$\times5$ dynamical matrix block. Matrix diagonalization leads then to four nonzero phonon frequencies corresponding to the four polar phonon modes, see Fig. \ref{phonons} (middle). As expected, the frequencies are independent of $\alpha$ leading to three-fold degeneracy of each mode.

For the AFM-I and AFM-II orderings the symmetry is lowered to tetragonal $D_{4h}$ and $D_{4}$ point groups, respectively, with the tetragonal direction chosen to be along the $z$ axis. In both cases the $T_{1u}$ representation becomes reducible resulting in a splitting of the triple degenerate polar phonon modes according to $T_{1u}\rightarrow A_{2u}\oplus E_u$ for AFM-I and $T_{1u}\rightarrow A_{2}\oplus E$ for AFM-II. The one-dimensional $A_{2u}$ and $A_2$ irreducible representations transform as a vector along the $z$ axis, while the two-dimensional $E_u$ and $E$ irreducible representations transform as a vector in the $xy$ plane.

In order to evaluate this splitting we calculated the polar phonons for the AFM-I and AFM-II orderings by diagonalizing the three 5$\times5$ blocks of the dynamical matrix using the $T_{1u}$ symmetry-adapted modes, $f_{n,\alpha}$ \cite{Comment}. Since we are interested in the splitting generated by the spin pattern alone, similarly as in Ref. \onlinecite{Massida} we neglected the magnetically induced tetragonal distortion of the crystal and performed calculations for the cubic structure found using FM configuration.

The results are shown in Fig. \ref{phonons}. The phonon frequencies obtained from the $\alpha=x$ and $\alpha=y$ dynamical matrix blocks are equal and form $E_u$ or $E$ doublets. On the other hand, the phonon frequencies obtained from the $\alpha=z$ block are different and correspond to $A_{2u}$ and $A_2$ singlets. The phonon splitting is the largest for the second lowest frequency mode (except for MgCr$_2$O$_4$ where it is the lowest frequency mode that has the highest splitting). In particular, for ZnCr$_2$O$_4$ in AFM-I state it becomes as large as 41 cm$^{-1}$. Interestingly, the magnitude of the splitting for the AFM-I order is, in general, about twice larger than for the AFM-II state.

The most important feature, however, is that for all considered compounds and for all the modes the singlet has a higher energy for the AFM-I state while for the AFM-II configuration it is the doublet that has a higher energy. This is in agreement with the experimentally observed sign reversal of the phonon splitting for ZnCr$_2$O$_4$ and MgCr$_2$O$_4$ compared with CdCr$_2$O$_4$\cite{Sushkov,Aguilar,Kant,Kant2} since the former ones have a ground state similar to AFM-I \cite{Tchernyshyov} while the magnetic ordering of the latter can be approximated by AFM-II \cite{Chern}.

\section{Spin-phonon coupling model}

In order to better understand the results of our first principles calculations we employ the spin-phonon coupling model in which the $T_{1u}$ block of the force-constant matrix for an arbitrary magnetic state is given by \cite{Fennie}

\begin{equation}
\tilde{C}_{n\alpha,n'\alpha'}=C_{n,n'}^{\text{PM}}+4\sum_{j}\frac{\partial^2J_{ij}}{\partial f_{n\alpha} \partial f_{n'\alpha'}}\langle\mathbf{S}_i\cdot\mathbf{S}_j\rangle
\end{equation}

Here, $\langle\mathbf{S}_i\cdot\mathbf{S}_j\rangle$ is the spin correlation function that is 1 for the FM ordering and either 1 or -1 for the AFM ordering. $C_{n,n'}^{\text{PM}}$ is the force-constants matrix in the PM phase. Note that the latter has a $O_h$ cubic symmetry and thus it doesn't depend on the $T_{1u}$ row indices $\alpha$ and $\alpha'$.

The above expression can be further simplified by using the symmetry of the magnetic state. In particular, the AFM-I and AFM-II (as well as FM) orderings don't induce couplings between different rows of $T_{1u}$ so the force-constant matrix is diagonal in the row indices: $\tilde{C}_{n\alpha,n'\alpha'}=\tilde{C}_{n\alpha,n'\alpha}\delta_{\alpha,\alpha'}\equiv \tilde{C}_{n,n'}(\alpha)$. Consequently, we only need $\partial^2J_{ij}/\partial f_{n\alpha} \partial f_{n'\alpha}$. Considering only the nn exchange interaction, there are only two types of such derivatives:\cite{Fennie} $J_{nn'\perp}''\equiv \partial^2J_{ij}/\partial f_{n\alpha} \partial f_{n'\alpha}$ $\forall$ $\mathbf{\hat r}_{ij}\cdot\mathbf{\hat \alpha}=0$ and $J_{nn'\parallel}''\equiv \partial^2J_{ij}/\partial f_{n\alpha} \partial f_{n'\alpha}$ $\forall$ $\mathbf{\hat r}_{ij}\cdot\mathbf{\hat \alpha}\neq0$, where $\mathbf{\hat r}_{ij}$ is the vector linking nn sites $i$ and $j$ and $\mathbf{\hat \alpha}$ is the vector along the $\alpha$ axis. Therefore, we can write

\begin{eqnarray}
\nonumber
 \tilde{C}_{n,n'}(\alpha)=C_{n,n'}^{\text{PM}}+4J_{nn'\perp}''\sum_{\mathbf{\hat r}_{\perp}}\langle\mathbf{S}_i\cdot\mathbf{S}_j\rangle \\
  +4J_{nn'\parallel}''\sum_{\mathbf{\hat r}_{\parallel}}\langle\mathbf{S}_i\cdot\mathbf{S}_j\rangle
\end{eqnarray}

where the first summation is over the two nn in the plane perpendicular to $\alpha$, and the second summation is over the other four nn. For the three magnetic orderings considered, we obtain

\begin{equation}
\tilde{C}^{\text{FM}}_{nn'}(\alpha=x,y,z)=C^{\text{PM}}_{nn'}+8J_{nn'\perp}''+16J_{nn'\parallel}''
\label{equ:asdf3}
\end{equation}

\begin{equation}
\tilde{C}^{\text{AFM-I}}_{nn'}(\alpha=x,y)=C^{\text{PM}}_{nn'}-8J_{nn'\perp}''
\label{equ:asdf2}
\end{equation}

\begin{equation}
\tilde{C}^{\text{AFM-I}}_{nn'}(\alpha=z)=C^{\text{PM}}_{nn'}+8J_{nn'\perp}''-16J_{nn'\parallel}''
\label{equ:asdf1}
\end{equation}

\begin{equation}
\tilde{C}^{\text{AFM-II}}_{nn'}(\alpha=x,y)=C^{\text{PM}}_{nn'}-8J_{nn'\parallel}''
\label{equ:asdf5}
\end{equation}

\begin{equation}
\tilde{C}^{\text{AFM-II}}_{nn'}(\alpha=z)=C^{\text{PM}}_{nn'}-8J_{nn'\perp}''
\label{equ:asdf4}
\end{equation}

The above equations explicitly demonstrate that in the FM state we have a three-fold degeneracy with respect to $\alpha$ and that in the AFM-I and AFM-II states these triplets split into a doublet ($\alpha=x,y$) and a singlet ($\alpha=z$).

The parameters $C^{\text{PM}}_{nn'}$, $J_{nn'\perp}''$, and $J_{nn'\parallel}''$ were fitted to the \emph{ab initio} force-constant matrices evaluated for FM, AFM-I, and AFM-II orderings. Essentially perfect fitting was obtained with the misfit lower than 0.03 meV/\AA\ and corresponding phonon frequencies within 1 cm$^{-1}$ of first principles values (see Supplementary Materials \cite{SM}). This indicates that the spin-phonon coupling model with only nn exchange coupling provides an excellent description of the effect of magnetic order on phonon frequencies in ACr$_2$O$_4$ spinels.

Explicit forms of $C^{\text{PM}}$, $J_{\perp}''$, and $J_{\parallel}''$ force-constant matrices for different ACr$_2$O$_4$ spinels are shown in Supplementary Materials \cite{SM}. We find that for all compounds $J_{33\perp}''$ is positive and significantly larger than any other element of $J_{\perp}''$ and $J_{\parallel}''$ matrices. As discussed in Ref. \onlinecite{Fennie}, the anomalously large value of $J_{33\perp}''$ originates from the exponential form of the direct exchange contribution ($J_d$) to $J_1$. Indeed, $J_d=Ae^{-aD_{Cr-Cr}}$ where $A$ and $a$ are positive constants and $D_{Cr-Cr}$ is the nn Cr-Cr bond length. The only partner function that significantly affects $D_{Cr-Cr}$ is $f_{3\alpha}$ (with $\hat\alpha$ perpendicular to the bond)\cite{Sushkov} resulting in large $J_{33\perp}''$. This explanation is consistent with the fact that the $J_{33\perp}''$ element is similar for MgCr$_2$O$_4$ and ZnCr$_2$O$_4$ compounds but it decreases with the size of A ion due to diminished role of the direct exchange mechanism. Note also that the positive sign of $J_{33\perp}''$ is the direct consequence of the exponential dependence of $J_d$ on the atomic displacements which requires the second derivative to have the same sign as $J_d$.

The second largest element among $J_{\perp}''$ and $J_{\parallel}''$ matrices is $J_{35\perp}''$ (or $J_{53\perp}''$). Since $f_{5\alpha}$ corresponds to the displacement of O sublattice which modifies Cr-O-Cr angle, this shows that the superexchange mechanism also contributes to the spin-phonon coupling. However, the superexchange contribution to the spin-phonon coupling is always significantly smaller than the direct exchange contribution. This remains true even for HgCr$_2$O$_4$ compound where the superexchange interaction is stronger than the direct exchange coupling. We believe that this relative ineffectiveness of the superexchange in generating a spin-phonon coupling is a generic feature and is due to the fact that this mechanism doesn't depend so strongly on atomic displacement as the direct exchange mechanism. This feature plays also an important role in the success of our spin-phonon coupling model where we considered only nn exchange interaction. Indeed, while for  for MgCr$_2$O$_4$ and ZnCr$_2$O$_4$ $J_1$ is at least order of magnitude larger than other exchange parameters, in the case of CdCr$_2$O$_4$ and HgCr$_2$O$_4$ compounds the $J_3$ and $J_3'$ couplings are not negligible and our approximation  works only because these couplings originates from superexchange processes and have weak dependence on atomic displacements.

Having established and understood the validity of our spin-phonon coupling we can now use it to provide an additional insight into magnetically induced phonon splitting. Since $J_{nn'\perp}''$ and $J_{nn'\parallel}''$ are much smaller that the elements of the paramagnetic force-constant matrix, the phonon frequencies can be written as

\begin{eqnarray}
\nonumber
 \tilde{\omega}_{\lambda}(\alpha)\approx \omega^{\text{PM}}_{\lambda}  + \\
 \frac{2}{\omega^{\text{PM}}_{\lambda}}
 \left(\mathcal{J}_{\lambda\perp}''\sum_{\mathbf{\hat r}_{\perp}}\langle\mathbf{S}_i\cdot\mathbf{S}_j\rangle
 +\mathcal{J}_{\lambda\parallel}''\sum_{\mathbf{\hat r}_{\parallel}}\langle\mathbf{S}_i\cdot\mathbf{S}_j\rangle\right)
\end{eqnarray}

Here, $\omega^{\text{PM}}_{\lambda}$ is the paramagnetic phonon frequency and for each phonon mode we introduced: $\mathcal{J}_{\lambda\perp,\parallel}''=u^{\dagger}_{\lambda}J_{\perp,\parallel}''u_{\lambda}$ where $u_\lambda$ are the paramagnetic dynamical matrix eigenvectors and $J_{\perp,\parallel}''$ is the dynamical matrix corresponding to $J_{nn'\perp,\parallel}''$ force-constant matrix. The magnetically-induced phonon splittings for the two principle AFM orders are then given by

\begin{eqnarray}
\nonumber
\Delta\omega_{\lambda}^{\text{AFM-I}}&\equiv&\tilde{\omega}^{\text{AFM-I}}_{\lambda}(\alpha=z)\ - \tilde{\omega}^{\text{AFM-I}}_{\lambda}(\alpha=x,y) \\
&\approx& 8\mathcal{J}_{\lambda\perp}''/\omega^{\text{PM}}_{\lambda} - 8\mathcal{J}_{\lambda\parallel}''/\omega^{\text{PM}}_{\lambda}
\label{delta1}
\end{eqnarray}

\begin{eqnarray}
\nonumber
\Delta\omega_{\lambda}^{\text{AFM-II}}&\equiv&\tilde{\omega}^{\text{AFM-II}}_{\lambda}(\alpha=z) - \tilde{\omega}^{\text{AFM-II}}_{\lambda} (\alpha=x,y) \\
&\approx& -4\mathcal{J}_{\lambda\perp}''/\omega^{\text{PM}}_{\lambda} + 4\mathcal{J}_{\lambda\parallel}''/\omega^{\text{PM}}_{\lambda}
\label{delta2}
\end{eqnarray}

The ratio $\mathcal{J}_{\lambda\perp,\parallel}''/\omega^{\text{PM}}_{\lambda}$ characterize the strength of the magnetic contribution  to the phonon frequencies. Table \ref{exchanges} shows these parameters for different ACr$_2$O$_4$ spinels. We can immediately observe that the only appreciable ratio are $\mathcal{J}_{\lambda\perp}''/\omega^{\text{PM}}_{\lambda}$ for the two lowest-frequency modes ($\lambda=$1,2) and both are always positive. It follows from Eqs. (\ref{delta1}) and (\ref{delta2}) that the phonon splitting is the largest for the two lowest-frequency modes and it is positive (negative) for the AFM-I (AFM-II) orderings. This feature is a direct consequence of $J_{33\perp}''$ being positive and dominant among other elements of $J_{\perp}''$ and $J_{\parallel}''$ since the $\lambda=$1,2 modes have the largest content of the $f_{3\alpha}$ partner function (see Supplementary Materials \cite{SM}). Therefore, according to the discussion above, different signs of the phonon splittings that we found from first principles for the lowest phonon modes are ultimately related to the dominant role of the direct exchange mechanism in generating the spin-phonon coupling.

Interestingly, $\mathcal{J}_{\lambda\perp}''/\omega^{\text{PM}}_{\lambda}$ for $\lambda=$3,4 are also always positive while $\mathcal{J}_{\lambda\parallel}''/\omega^{\text{PM}}_{\lambda}$ for all modes are consistently negative. This results in the sign of the phonon splitting to be positive (negative) for the AFM-I (AFM-II) states for the two-highest phonon modes as well. These features, however, is difficult to explain microscopically due to small values of the splittings. In fact, the spliting of the two-highest phonon modes is too small to be seen in experiments.

According to Eqs. (\ref{delta1}) and (\ref{delta2}) we can write

\begin{equation}
\Delta\omega_{\lambda}^{\text{AFM-II}}\approx-\frac{1}{2}\Delta\omega_{\lambda}^{\text{AFM-I}}
\label{deltas}
\end{equation}

Therefore, the phonon splittings for AFM-I and AFM-II orders are always opposite and the latter is approximately half of the former. This is a general result which follows directly from the validity of the nn spin-phonon coupling model and it is independent on the signs and sizes of the exchange second derivatives. Note that the above relation is well satisfied by first principles data (Fig. \ref{phonons}) which again demonstrates validity of the model. Note, however, that this relation doesn't tell us for which ordering the phonon splitting is positive. In order to answer this question more microscopic analysis (as above) is needed.

\section{Conclusions}

In summary, we investigated the effect of magnetic ordering on phonon frequencies of ACr$_2$O$_4$ spinels using first prinicples electronic structure calculations. We found that our \emph{ab initio} results are very well described by the spin-phonon coupling model with only nn exchange coupling \cite{Fennie}. Both the model and first principles calculations show that a specific type of spin ordering has a crucial effect on magnetically induced phonon splitting. In particular, we found that the different magnetic states observed in different spinels lead to the opposite signs of the phonon splittings observed in ZnCr$_2$O$_4$ and MgCr$_2$O$_4$ compounds compared to CdCr$_2$O$_4$. This feature is a result of an important role played by the direct exchange mechanism in generating the spin-phonon coupling in these materials.

%==================================================
\acknowledgements
%==================================================
We acknowledge fruitful discussions with Craig J. Fennie and Karin Rabe. Work at Ames Lab was supported by the U.S. Department of Energy (DOE), Office of Science, Basic Energy Sciences, Materials Science and Engineering Division. Ames Laboratory is operated for the U.S. DOE by Iowa State University under contract \#DE-AC02-07CH11358. T.B. was supported by the Rutgers Center for Materials Theory.

\pagebreak
\begin{widetext}
\begin{center}
\textbf{\large Supplementary Materials for "Magnetically-induced phonon splitting in ACr$_2$O$_4$ spinels from first principles"}
\end{center}
\end{widetext}

\setcounter{equation}{0}
\setcounter{figure}{0}
\setcounter{table}{0}
\setcounter{section}{0}
\setcounter{page}{1}

\makeatletter
\renewcommand{\theequation}{S\arabic{equation}}
\renewcommand{\thefigure}{S\arabic{figure}}
\renewcommand{\bibnumfmt}[1]{[S#1]}
\renewcommand{\citenumfont}[1]{S#1}

\section{Structural data}

\begin{table}[H]
\centering{}
\begin{tabular}{r|c|c|c|c}
\hline
\hline
Compound            & $a_0$ (\AA)& $x$ & $D_{Cr-Cr}$ (\AA)& $\alpha$ ($^0$)\\
\hline
MgCr$_2$O$_4$	& 8.372	&0.262 &  2.960 &95.4\\
ZnCr$_2$O$_4$ 	& 8.351	&0.262 &  2.952 &95.4\\
CdCr$_2$O$_4$	& 8.628	&0.269 & 3.051  & 98.8\\
HgCr$_2$O$_4$	& 8.703	&0.271 & 3.077  & 99.8 \\
\hline
\hline
\end{tabular}
\caption{Structural parameters of different ACr$_2$O$_4$ spinels in the $Fd\bar3m$ cubic phase calculated for the ferromagnetic spin configuration. In addition to the lattice parameter ($a_0$) and the fractional coordinate of oxygen atoms ($x$) we also show the nearest-neighbor Cr-Cr distance ($D_{Cr-Cr}$) and the Cr-O-Cr angle ($\alpha$).}
\label{lattice}
\end{table}

\section{Symmetry adapted modes}

\begin{figure*}
\begin{center}
\includegraphics[angle=270,width=1.0\hsize]{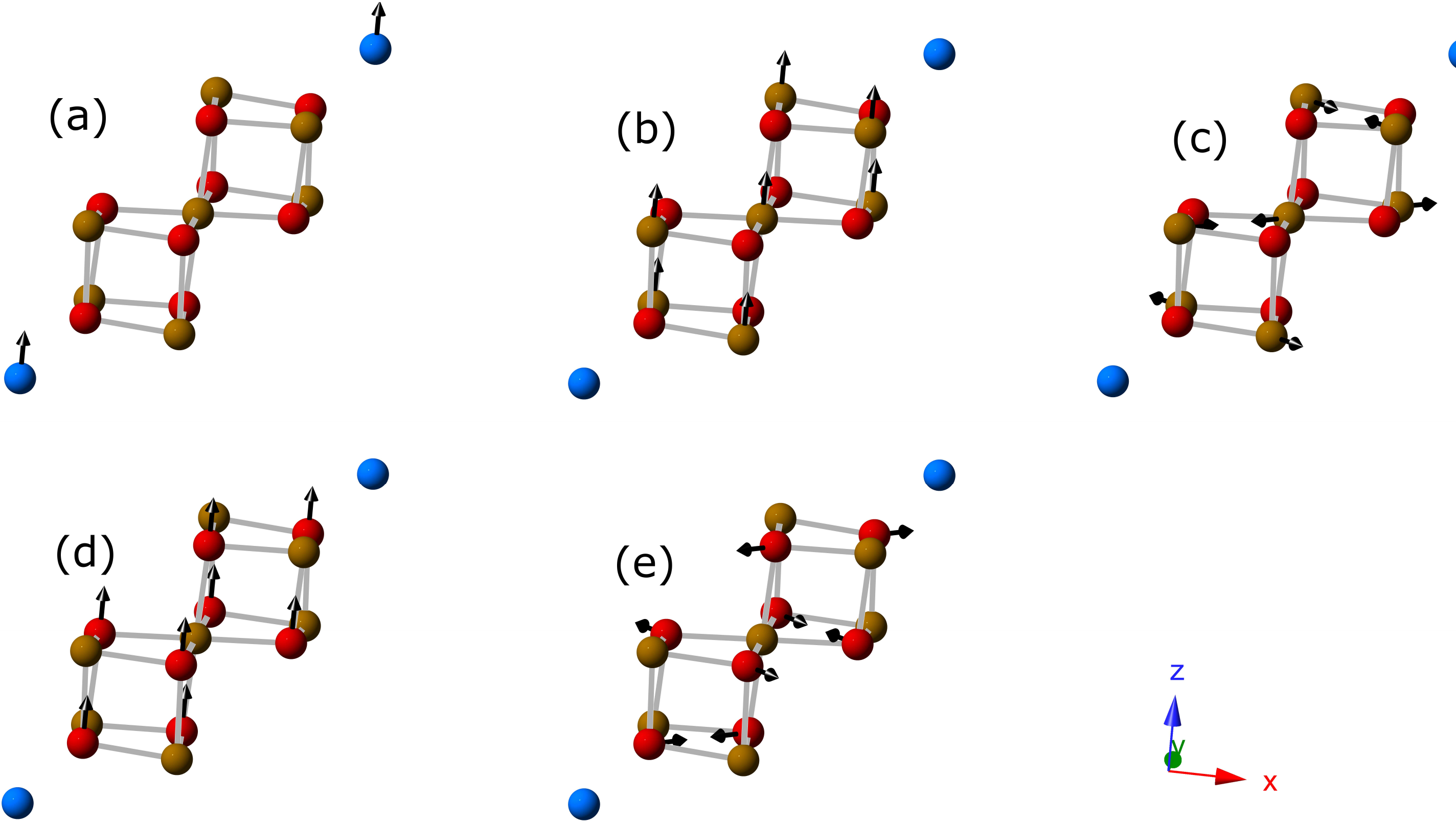}
\end{center}
\caption{Symmetry-adapted modes $f_{n,z}$ ($n=1,2,3,4,5$): (a) $f_{1,z}$, (b) $f_{2,z}$, (c) $f_{3,z}$, (d) $f_{4,z}$, (e) $f_{5,z}$. Blue, brown, and red spheres denote A, Cr, and O atoms, respectively while the arrows indicate atomic displacements. Symmetry-adapted modes  $f_{n,x}$  ($f_{n,y}$) can be obtained by the 90$^{\circ}$ rotation around the $y$ ($x$) axis.}
\label{modes}
\end{figure*}

The symmetry-adapted modes $f_{n,z}$ ($n=1,2,3,4,5$) are illustrated in Fig. \ref{modes}. The $f_{n,x}$ and $f_{n,y}$ modes can be obtained from $f_{n,z}$ by the 90$^{\circ}$ rotation around the $y$ and the $x$ axis, respectively.

\section{Fitting results}

Fitted values of the parameters of the spin-phonon coupling model for different ACr$_2$O$_4$ compounds are shown below. The units are meV/\AA.

\textbf{MgCr$_2$O$_4$}

\begin{equation}
C^{\text{PM}}=
\left(
\begin{array}{ccccc}
  10.464  &  -2.471  &  -3.373  &  -3.485  &  -4.601 \\
  -2.469  &  20.248  &   5.114  &  -13.082  &  -1.649 \\
  -3.372  &   5.115  &  30.768  &  -1.931  &  -9.617 \\
  -3.488  &  -13.072  &  -1.930  &  10.987  &   3.467 \\
  -4.607  &  -1.645  &  -9.602  &   3.467  &  16.057 \\
\end{array}
\right)
\end{equation}

\begin{equation}
J_{\perp}''=
\left(
\begin{array}{rrrrr}
  -0.001  &  -0.000  &  -0.008  &   0.000  &  -0.001 \\
  -0.000  &   0.036  &   0.021  &  -0.025  &  -0.002 \\
  -0.008  &   0.021  &   0.293  &  -0.011  &   0.103 \\
   0.000  &  -0.025  &  -0.011  &   0.017  &   0.002 \\
  -0.001  &  -0.003  &   0.103  &   0.002  &   0.029 \\
\end{array}
\right)
\end{equation}

\begin{equation}
J_{\parallel}''=
\left(
\begin{array}{rrrrr}
   0.000  &  -0.004  &  -0.002  &   0.002  &   0.001 \\
  -0.003  &  -0.039  &  -0.004  &   0.029  &   0.006 \\
  -0.002  &  -0.004  &  -0.024  &   0.004  &   0.010 \\
   0.002  &   0.029  &   0.004  &  -0.022  &  -0.005 \\
   0.001  &   0.006  &   0.011  &  -0.005  &  -0.007 \\
\end{array}
\right)
\end{equation}

\textbf{ZnCr$_2$O$_4$}

\begin{equation}
C^{\text{PM}}=
\left(
\begin{array}{ccccc}
   9.269  &  -2.470  &  -4.047  &  -2.888  &  -3.815 \\
  -2.471  &  20.804  &   4.428  &  -13.475  &  -1.104 \\
  -4.049  &   4.426  &  31.416  &  -1.106  &  -10.057 \\
  -2.888  &  -13.467  &  -1.108  &  10.967  &   2.690 \\
  -3.817  &  -1.105  &  -10.049  &   2.690  &  16.631 \\
\end{array}
\right)
\end{equation}

\begin{equation}
J_{\perp}''=
\left(
\begin{array}{rrrrr}
  -0.000  &   0.002  &  -0.012  &  -0.001  &  -0.001 \\
   0.002  &   0.042  &   0.019  &  -0.031  &   0.002 \\
  -0.011  &   0.019  &   0.298  &  -0.008  &   0.105 \\
  -0.002  &  -0.031  &  -0.008  &   0.023  &  -0.001 \\
  -0.001  &   0.002  &   0.106  &  -0.001  &   0.027 \\
\end{array}
\right)
\end{equation}

\begin{equation}
J_{\parallel}''=
\left(
\begin{array}{rrrrr}
  -0.000  &  -0.004  &  -0.001  &   0.003  &   0.002 \\
  -0.004  &  -0.040  &  -0.005  &   0.030  &   0.007 \\
  -0.001  &  -0.005  &  -0.027  &   0.004  &   0.012 \\
   0.003  &   0.030  &   0.004  &  -0.023  &  -0.006 \\
   0.002  &   0.007  &   0.012  &  -0.006  &  -0.009 \\
\end{array}
\right)
\end{equation}

\textbf{CdCr$_2$O$_4$}

\begin{equation}
C^{\text{PM}}=
\left(
\begin{array}{ccccc}
   9.969  &  -2.286  &  -3.714  &  -3.368  &  -5.921 \\
  -2.286  &  18.870  &   4.419  &  -12.200  &  -0.692 \\
  -3.711  &   4.417  &  28.057  &  -1.268  &  -7.973 \\
  -3.370  &  -12.190  &  -1.270  &  10.305  &   3.449 \\
  -5.924  &  -0.692  &  -7.962  &   3.451  &  16.765 \\
\end{array}
\right)
\end{equation}

\begin{equation}
J_{\perp}''=
\left(
\begin{array}{rrrrr}
  -0.000  &   0.002  &  -0.006  &  -0.001  &  -0.002 \\
   0.002  &   0.035  &   0.034  &  -0.026  &  -0.001 \\
  -0.006  &   0.034  &   0.169  &  -0.021  &   0.078 \\
  -0.002  &  -0.026  &  -0.021  &   0.019  &   0.001 \\
  -0.002  &  -0.001  &   0.078  &   0.002  &   0.044 \\
\end{array}
\right)
\end{equation}

\begin{equation}
J_{\parallel}''=
\left(
\begin{array}{rrrrr}
  -0.000  &  -0.002  &  -0.000  &   0.001  &   0.001 \\
  -0.002  &  -0.019  &   0.002  &   0.015  &   0.000 \\
  -0.001  &   0.002  &  -0.008  &  -0.001  &   0.004 \\
   0.001  &   0.015  &  -0.001  &  -0.011  &  -0.000 \\
   0.001  &   0.000  &   0.004  &  -0.000  &  -0.001 \\
\end{array}
\right)
\end{equation}

\textbf{HgCr$_2$O$_4$}

\begin{equation}
C^{\text{PM}}=
\left(
\begin{array}{ccccc}
   7.196  &  -2.278  &  -4.123  &  -1.987  &  -4.854 \\
  -2.283  &  18.472  &   3.903  &  -11.920  &  -0.190 \\
  -4.127  &   3.909  &  27.101  &  -0.700  &  -7.537 \\
  -1.987  &  -11.913  &  -0.700  &   9.418  &   2.562 \\
  -4.854  &  -0.197  &  -7.528  &   2.566  &  16.317 \\
\end{array}
\right)
\end{equation}

\begin{equation}
J_{\perp}''=
\left(
\begin{array}{rrrrr}
  -0.000  &   0.004  &  -0.007  &  -0.003  &  -0.004 \\
   0.004  &   0.036  &   0.034  &  -0.027  &   0.004 \\
  -0.007  &   0.034  &   0.140  &  -0.021  &   0.071 \\
  -0.003  &  -0.027  &  -0.021  &   0.021  &  -0.001 \\
  -0.004  &   0.004  &   0.070  &  -0.001  &   0.042 \\
\end{array}
\right)
\end{equation}

\begin{equation}
J_{\parallel}''=
\left(
\begin{array}{rrrrr}
  -0.000  &  -0.002  &  -0.001  &   0.002  &   0.001 \\
  -0.002  &  -0.016  &   0.003  &   0.013  &  -0.000 \\
  -0.000  &   0.003  &  -0.006  &  -0.002  &   0.003 \\
   0.002  &   0.013  &  -0.002  &  -0.010  &  -0.000 \\
   0.001  &  -0.000  &   0.003  &  -0.000  &  -0.001 \\
\end{array}
\right)
\end{equation}

The phonon frequencies for FM, AFM-I, and AFM-II orderings evaluated from the spin-phonon coupling model are shown in Table \ref{freqs}. As expected, the results agree very well with first principles calculations (see Fig. 3 of the main manuscript). In addition, we show the paramagnetic phonon frequencies.

\begin{table}[H]
\centering{}
\begin{tabular}{c|c|c|c|c|c|c|c}
\hline
\hline
Compound           & $\lambda$ & \multicolumn{6}{|c} {$\omega_\lambda$} \\
\cline{3-8}
                            &                   & FM & \multicolumn{2}{|c} {AFM-I} & \multicolumn{2}{|c} {AFM-II} & PM  \\
\cline{3-8}
                            &                   & $x,y,z$ & $x,y$  & $z$                  &   $x,y$  & $z$                       & $x,y,z$   \\
\hline
MgCr$_2$O$_4$	& 1 & 261 & 224 & 261 & 245 & 224 & 245  \\
\cline{2-8}
                        	& 2 & 424 & 406 & 425 & 415 & 406 & 415  \\
\cline{2-8}
                        	& 3 & 477 & 479 & 488 & 484 & 479 & 481  \\
\cline{2-8}
                        	& 4 & 612 & 615 & 623 & 619 & 615 & 616  \\
\hline
ZnCr$_2$O$_4$	& 1 & 189 & 173 & 189 & 182 & 173 & 182  \\
\cline{2-8}
                        	& 2 & 379 & 340 & 381 & 360 & 340 & 360  \\
\cline{2-8}
                        	& 3 & 499 & 499 & 511 & 505 & 499 & 502  \\
\cline{2-8}
                        	& 4 & 599 & 603 & 611 & 607 & 603 & 604  \\
\hline
CdCr$_2$O$_4$	& 1 & 151 & 141 & 151 & 146 & 141 & 146  \\
\cline{2-8}
                        	& 2 & 369 & 343 & 370 & 356 & 343 & 356  \\
\cline{2-8}
                        	& 3 & 473 & 471 & 481 & 476 & 471 & 474  \\
\cline{2-8}
                        	& 4 & 598 & 594 & 601 & 597 & 594 & 596  \\
\hline
HgCr$_2$O$_4$	& 1 & 108 & 101 & 108 & 105 & 101 & 105  \\
\cline{2-8}
                        	& 2 & 357 & 332 & 358 & 345 & 332 & 345  \\
\cline{2-8}
                        	& 3 & 475 & 471 & 483 & 477 & 471 & 475  \\
\cline{2-8}
                        	& 4 & 575 & 571 & 577 & 574 & 571 & 573  \\
\hline
\hline
\end{tabular}
\caption{Phonon frequencies for different magnetic orderings calculated from the spin-phonon coupling model. The units are cm$^{-1}$.}
\label{freqs}
\end{table}

\section{Paramagnetic dynamical matrix eigenvectors}

\begin{table}[H]
\centering{}
\begin{tabular}{c|c|c|c|c|c|c}
\hline
\hline
Compound      & $\lambda$ & \multicolumn{5}{|c} {$n$} \\
\cline{3-7}
                            &                   & 1 & 2 & 3 & 4 & 5 \\
\hline
MgCr$_2$O$_4$               & 1 & 0.38 & 0.05 & 0.36 & 0.01 & 0.20 \\
\cline{2-7}
                        	& 2 & 0.45 & 0.06 & 0.45 & 0.01 & 0.03 \\
\cline{2-7}
                        	& 3 & 0.00 & 0.27 & 0.07 & 0.42 & 0.24 \\
\cline{2-7}
                        	& 4 & 0.04 & 0.08 & 0.12 & 0.23 & 0.52 \\
\hline
ZnCr$_2$O$_4$               & 1 & 0.58 & 0.16 & 0.12 & 0.07 & 0.07 \\
\cline{2-7}
                        	& 2 & 0.13 & 0.04 & 0.65 & 0.01 & 0.16 \\
\cline{2-7}
                        	& 3 & 0.00 & 0.28 & 0.06 & 0.47 & 0.19 \\
\cline{2-7}
                        	& 4 & 0.01 & 0.07 & 0.16 & 0.18 & 0.58 \\
\hline
CdCr$_2$O$_4$               & 1 & 0.51 & 0.24 & 0.08 & 0.11 & 0.06 \\
\cline{2-7}
                        	& 2 & 0.07 & 0.03 & 0.78 & 0.02 & 0.10 \\
\cline{2-7}
                        	& 3 & 0.00 & 0.30 & 0.03 & 0.44 & 0.22 \\
\cline{2-7}
                        	& 4 & 0.01 & 0.06 & 0.11 & 0.20 & 0.62 \\
\hline
HgCr$_2$O$_4$               & 1 & 0.42 & 0.32 & 0.05 & 0.18 & 0.03 \\
\cline{2-7}
                        	& 2 & 0.03 & 0.02 & 0.81 & 0.02 & 0.12 \\
\cline{2-7}
                        	& 3 & 0.00 & 0.33 & 0.02 & 0.48 & 0.16 \\
\cline{2-7}
                        	& 4 & 0.00 & 0.05 & 0.12 & 0.14 & 0.69 \\
\hline
\hline
\end{tabular}
\caption{Squared projections of the paramagnetic dynamical matrix eigenvectors, $u_{\lambda}$, on the symmetry adapted partner functions, $f_{n\alpha}$.}
\label{eigens}
\end{table}

\begin{thebibliography}{36}
\bibitem [{Com()}]{Comment}%
  \BibitemOpen
  \href@noop {} {\bibinfo  {journal} {Under $D_{4h}$ or $D_4$ certain rows of
  the silent $T_{2u}$ irreducible representation of $O_h$ become ir-active and,
  can in principle mix with the irreducible representations originating from
  $T_{1u}$. This mixing is however small (as we explicitly checked) and
  therefore was neglected}\ }\BibitemShut {NoStop}%
\end{thebibliography}
\end{document}